\documentstyle[12pt]{article}
\begin{document}
\title
{ Model of graviton-dusty universe}
\author
{By Michael A. Ivanov \\
Chair of Physics, \\
Belarus State University of Informatics and Radioelectronics, \\
6 P. Brovka Street,  BY 220027, Minsk, Republic of Belarus.\\
E-mail: ivanovma@gw.bsuir.unibel.by.}
\date{June 10, 2002}
\maketitle

\begin{abstract}
Primary features of a new cosmological model, which is based on
conjectures about an existence of the graviton background and
superstrong gravitational quantum interaction, are considered. An
expansion of the universe is impossible in such the model because of
deceleration of massive objects by the graviton background, which is
similar to the one for the NASA deep space probes Pioneer 10, 11.
Redshifts of remote objects are caused in the model by interaction
of photons with the graviton background, and the Hubble constant
depends on an intensity of interaction and an equivalent temperature
of the graviton background. Virtual massive gravitons would be dark
matter particles. They transfer energy, lost by luminous matter
radiation, which in a final stage may be collected with black holes
and other massive objects.
\end{abstract}

\section[1]{Introduction }

In present cosmological models, based on the big bang conjecture,
redshifts of remote objects are explained as the Doppler effect.
Observation of small additional acceleration of NASA deep space
probes \cite{15,40,25} may be interpreted in such a manner, that
redshifts of galaxies turn out a quantum gravity effect. As well as
probes' acceleration, it would be caused by a hypothetical
superstrong gravitational quantum interaction with the graviton
background \cite{41}. This new hypothesis about the redshift nature
finds further confirmation in confrontation \cite {41} with
Supernova Search Team data \cite{42}.
\par
In this paper, some features of the cosmological model, which may be
constructed on a base of such a new approach, are discussed. They
are: the redshift nature; a transfer of energy, which is lost by
luminous matter radiation, with virtual massive gravitons; an
increment of lifetime of such gravitons due to sequent decreasing
their energy by collisions with gravitons of the background; final
utilization of virtual massive gravitons in black holes and other
massive objects.
\par

\section[2]{Hypothetical superstrong gravitational quantum
interaction and its consequences}

If the graviton background exists with the equivalent temperature
upwards $3K$, then collisions of photons with gravitons of this
background will lead to photon redshifts if the interaction is
strong enough. A photon energy $E(r)$ will change with increasing of
a distance $r$ from a source as
\begin{equation}
                      E(r)=E_{0} \exp(-ar),
\end{equation}
where $E_{0}$ is an initial value of energy, $a=H/c,$ $H$ is the
Hubble constant.
\par
It is shown in author's paper \cite{41}, that a cross-section
$\sigma (E,\omega)$ of interaction of a photon with an energy $E$
with a graviton, having an energy $\omega,$ as well as the Hubble
constant $H$ may be expressed with the help of a new dimensional
constant $D:$
\begin{equation}
\sigma (E,\omega)= D \cdot E \cdot \omega,
\end{equation}
\begin{equation}
H= {1 \over 2\pi} D \cdot \bar \omega \cdot (\sigma T^{4}),
\end{equation}
where $\bar \omega$ is an average graviton energy, $\sigma$ is the
Stephan-Boltzmann constant, with $D \sim 10^{-27} m^{2}/eV^{2}$ if
whole redshift is caused by such the interaction.
\par
An universality of gravitational interaction means that any massive
body, moving relatively to the background with a velocity $v$, must
feel a deceleration $w,$ which is equal to:
\begin{equation}
w = - Hc(1-v^{2}/c^{2}).
\end{equation}
For $Hc=(4.8 - 7.2) \cdot 10^{-10} m/s^{2}$ by $H=(1.6 - 2.4) \cdot
10^{-18} s^{-1}$ (i.e. by $H=(50 - 75) km \cdot s^{-1} \cdot Mpc^{-
1}),$ a magnitude of this deceleration corresponds to the one of
observed additional acceleration for Pioneer 10 \cite{15,40,25}. The
acceleration $w$ should be directed against a body velocity
relatively to the graviton background. The refined data \cite {40}
show annual periodic variations of the apparent acceleration.
Perhaps, such the variations are connected with Earth's additional
acceleration under its orbital motion due to the same interaction
with the graviton background \cite{41a}.
\par
If redshifts of galaxies are really caused by such the effect, then
an expected picture of visible Universe changes. A region of the
Universe, which is visible by an observer, will not be bounded with
a sphere of the Hubble radius $R_{0}=c/H,$ but any source with a
temperature $T_{s}$ may be picked out by an observer above the
microwave background on the distance
\begin{equation}
R < R_{0} \ln \frac { T_{s}}{T},
\end{equation}
i.e. for a source with $T_{s} \simeq 6000 K$ we have $ R < 7.6
R_{0}.$ It is $ R < (100 - 150)$ Gyr for $R_{0} \simeq (13.5 - 20)$
Gyr. An estimate of distances to objects with given $z$ is changed too;
for example, the quasar with $z=5.8$ \cite {43} should be in a
distance approximately twice bigger than the one expected in the
model based on the Doppler effect.

\section[3]{Utilization of energy, which is lost by visible matter radiation}

Unlike models of expanding universe, in any tired light model one
has a problem of utilization of energy, lost by radiation of remote
objects. In the model, a virtual graviton forms under collision of a
photon with a graviton of the graviton background. It should be
massive if an initial graviton transfers its total momentum to a
photon; it follows from the energy conservation law that its energy
$\omega^{'}$ must be equal to $2 \omega$ if $\omega$ is an initial
graviton energy. In force of the uncertainty relation, one has for a
virtual graviton lifetime $\tau:$ $\tau \leq \frac
{\hbar}{\omega^{'}},$ i.e. for $\omega^{'} \sim 10^{-4} eV$ it is
$\tau \leq 10^{-11} s.$ In force of conservation laws for energy,
momentum and angular momentum, a virtual graviton may decay into no
less than three real gravitons. In a case of decay into three
gravitons, its energies should be equal to $\omega, \omega^{''},
\omega {'''},$ with $\omega^{''} + \omega {'''}= \omega.$ So, after
this decay, two new gravitons with $\omega^{''}, \omega {'''} <
\omega$ inflow into the graviton background. It is a source of
adjunction of the graviton background.
\par
From another side, an interaction of gravitons of the background
between themselves should lead to the formation of virtual massive
gravitons too, with energies less than $\omega_{min}$ where
$\omega_{min}$ is a minimal energy of one graviton of an initial
interacting pare. If gravitons with energies $\omega^{''}, \omega
{'''}$ wear out a file of collisions with gravitons of the
background, its lifetime has increased. In every such a cycle
collision-decay, an average energy of "redundant" gravitons will
double decrease, and its lifetime will double increase. Only for
$\sim 93$ cycles, a lifetime will have increased from $10^{-11} s$ to
$10$ Gyr. Such virtual massive gravitons, with a lifetime increasing
from one collision to another, would duly serve dark matter
particles. Having a zero (or near to zero) initial velocity
relatively to the graviton background, the ones will not interact
with matter in any manner except usual gravitation. An ultracold gas
of such gravitons will condense under influence of gravitational
attraction into black holes or other massive objects. Additionally
to it, even in absence of initial heterogeneity, the one will easy
arise in such the gas that would lead to arising of super compact
massive objects, which will be able to turn out "germs" of black
holes. It is a method "to cold" the graviton background.
\par
So, the graviton background may turn up "a perpetual engine" of the
Universe, pumping energy from radiation to massive objects. An
equilibrium state of the background will be ensured by such the
temperature $T,$ for which an energy profit of the background due to
an influx of energy from radiation will be equal to a loss of its
energy due to a catch of virtual massive gravitons with black holes
or other massive objects. In such the picture, the chances are that
black holes should turn out "germs" of galaxies. After accumulation
of big enough energy by a black hole (to be more exact, by a super
compact massive object) by means of a catch of virtual massive
gravitons, the one must be absolved from an energy excess in via
ejection of matter, from which stars of galaxy should form. It
awaits to understand else in such the model how usual matter
particles form from virtual massive gravitons. It is optimistic that
the model of two-component fundamental fermions by the author \cite{46}
owns all symmetries of the standard model of elementary
particles (on global level). Perhaps, virtual gravitons with very
small masses are fully acceptable to the role of components of such
the system. Observation of non-zero neutrino mass \cite{45}
increases chances of this model of the fundamental fermions since
there is an additional right singlet in the model, which is able to
provide a non-zero neutrino mass. Chances of the model will rise
still more, if any particle of the forth generation or some indirect
indication of its existence will be detected. In author's paper
\cite{44}, the model of gravity in flat 12-space was described with
global $U(1)-$symmetry, in which a possibility exists to introduce
$SU(2)-$symmetry. I hope that unification of these models may
give us a clue to hidden still unity of gravity and other known
interactions.
\par

\section[4]{Conclusion}

Observations of last years give us strong evidences for supermassive
black holes in active and normal galactic nuclei \cite{5,7,8,12,16}
(of course, a central dark mass in galactic nucleus may not be a
black hole, but it is most likely to the one by its properties from
all known objects; one must remember that we know only that these
objects are supermassive and compact). The available evidence is
consistent with a suggestion that a majority of galaxies has black
holes \cite{5,3}. The discovery by Gebhardt et al. \cite{14} and
Ferrarese and Merritt \cite{10} of a correlation between nuclear
black hole mass and stellar velocity dispersion in elliptical
galaxies and spiral bulges shows that black holes are "native" for
host galaxies. Massive nuclear black holes of $10^{6} - 10^{9}$
solar masses may be responsible for the energy production in quasars
and active galaxies \cite{5}. Doppler-shifted emission lines in the
spectrum of active galactic nuclei are likely to originate from
relativistic outflows (or jets) in the vicinity of the central black
hole \cite{9}. Black hole candidates are also known in binaries,
supernovae, and clusters.
\par
In a frame of the model \cite{49} it was suggested that central
black holes of early-type galaxies grew adiabatically in homogeneous
isothermal cores due to matter accretion. In the present model, a
role of black holes in evolution of the Universe is changed; the
ones may be collectors of virtual massive gravitons and "germs" of
galaxies. Additionally, the growth of black hole mass inside of
future supernova stars would lead to their instability and formation
of supernovae.
\par
In author's papers \cite{47,48,41}, the methods were considered how to
verify the conjecture about the described non-dopplerian nature of
redshifts. One of them is a ground-based experiment with a
superstable laser radiation: if the conjecture is true, then a laser
radiation frequency after a delay line should be red shifted too. I
believe, that creation of necessary superstable lasers with
instability $\sim 10^{-17}$ would be speeded up after perception by
the scientific community of importance of such the verification.


\end{document}